\documentclass[%
 reprint,
 amsmath,amssymb,
 aps,
longbibliography
]{revtex4-1}

\usepackage{graphicx}
\usepackage{dcolumn}
\usepackage{bm}

\usepackage[colorlinks=true,citecolor=blue,linkcolor=blue, urlcolor=cyan]{hyperref} 

\usepackage{empheq}
\usepackage{color}
\usepackage[normalem]{ulem}
\usepackage[english]{babel}

\renewcommand{\Im}{\mathrm{Im}}

\newcommand{\eps}{\varepsilon}
\newcommand{\h}{\hbar}

\newcommand{\stkout}[1]{\ifmmode\text{\sout{\ensuremath{#1}}}\else\sout{#1}\fi}

\begin{document}

\title{Longitudinal-transverse splitting and fine structure of Fermi polarons\\ in two-dimensional semiconductors}

\author{Z.A. Iakovlev}
\affiliation{Ioffe Institute, 194021 St. Petersburg, Russia}

\author{M.M. Glazov}
\affiliation{Ioffe Institute, 194021 St. Petersburg, Russia}

\begin{abstract}
Interaction of excitons with resident charge carriers in semiconductors gives rise to bound three-particle complexes, trions, whose optical response is conveniently described in the framework of many-body correlated Fermi polaron states. These states are formed as a result of correlation of photocreated trion with the Fermi sea hole and possess the angular momentum component of $\pm 1$ depending on the helicity of the photon. We study theoretically the energy spectrum fine structure of Fermi polarons in  two-dimensional semiconductors based on transition metal dichalcogenides. We demonstrate both by the symmetry analysis and microscopic calculation  that the Fermi polarons with nonzero in-plane wavevector $\bm k$ are split, similarly to the neutral exciton states, into the linearly polarized longitudinal and transverse, with respect to the $\bm k$,  states. The origin of this longitudinal-transverse splitting is the long-range electron-hole exchange interaction that can be also described as the interaction of Fermi polarons with their induced  electromagnetic field. The effective Hamiltonian describing the Fermi polaron fine structure is derived, and its parameters are determined from the microscopic model.

\emph{Submitted to a special issue of J. Lumin. in homage to Profs.  F. Auzel and  A. Kaplyanskii.}
\end{abstract}

\maketitle

\section{Introduction}\label{sec:intro}

Optical properties of semiconductors and semiconductor nanostructures are mainly controlled by excitonic species: neutral excitons, Coulomb-bound electron-hole pairs, excitonic molecules or biexcitons, and charged excitons or trions~\cite{ivchenko05a,Klingshirn2012}. These quasiparticles usually determine absorption, reflection of light and underlie luminescence. Two-dimensional (2D) transition metal dichalcogenides (TMDC) described by the basic formula MX$_2$ where M stands for a transition metal, usually, Mo or W, and X for a chalcogen, X~=~S, Se, and Te, are direct band gap semiconductors with outstanding optical response dominated by tightly bound excitons and trions~\cite{Splendiani:2010a,Mak:2010bh,Mak:2013lh,Chernikov:2014a}, see Refs.~\cite{Kolobov2016book,RevModPhys.90.021001,Durnev_2018,Schneider:2018aa,Tartakovskii:2020aa,Glazov_2021} for reviews. 

It is not surprising that classical topics of optical spectroscopy of semiconductors are studied in 2D TMDCs. For example, upconversion effects or summation of excitation quanta discovered back in the days by Prof. Fran\c{c}ois Auzel~\cite{auzel1,auzel2,Auzel:2020aa} [see also related and independent works by V.V. Ovsyankin and P.P. Feofilov~\cite{ovsyankin1966mechanism,Feofilov:67}] are observed and actively studied in TMDC monolayers (MLs)~\cite{Jones:2016aa,Manca:2017aa,PhysRevX.8.031073,Jadczak:2019aa,Jadczak:2021aa}. Similarly, the excitation of valleys by polarized light~\cite{Mak:2012qf,Kioseoglou,PhysRevLett.112.047401,Tang:2019ab,Robert:2021wc},  spectroscopy of defects~\cite{Barja:2019aa,PhysRevLett.123.076801,Mitterreiter:2021aa}, and exciton-phonon interaction~\cite{PhysRevLett.119.187402,shree2018exciton,Brem:2020ud} in 2D TMDCs are in focus of research, these are the topics pursued by Prof. Alexander Kaplyanskii for classical semiconductors~\cite{Feofilov:1962aa,KAPLYANSKII197627,Gastev1982}. 

Needless to say that basic physical principles whose foundations were laid in seminal works of Auzel and Kaplyanskii remain the same, however, particular manifestations of the effects and their specific features turn out to be qualitatively and quantitatively different in 2D semiconductors. It makes optical properties of two-dimensional materials a vibrant and rapidly developing field of research. Key novel features of 2D TMDCs are (i) the presence of two valleys $\bm K_+$ and $\bm K_-$ at the Brillouin zone edges where the optical transitions are excited by the photons of opposite helicity, respectively, $\sigma^+$ and $\sigma^-$ and (ii) strong Coulomb interaction that gives rise to a plethora of many-body exciton-based states. 

Here we develop the theory of the energy spectrum fine structure of Fermi polarons, also known as Suris tetrons~\cite{PhysRevLett.112.147402}, the correlated complexes of a trion -- charged exciton -- and a hole in the Fermi sea of the resident charge carriers. Such effectively four particle bound states formed of the electron and hole in the exciton, resident electron picked out from the Fermi sea to form a trion, and the Fermi sea hole generated as a result of the Fermi sea excitation govern the optical response of doped 2D semiconductors~\cite{PSSB:PSSB343,suris:correlation,Sidler:2016aa,PhysRevB.95.035417,PhysRevB.102.085304,tiene2022crossover}. In several important cases like the redistribution of the oscillator strength~\cite{astakhov00,Courtade:2017a}, recoil effects in optical emission~\cite{PhysRevB.105.075311}, and diffusion at low carrier density~\cite{wagner:trions} the trion and Fermi polaron descriptions provide essentially the same results~\cite{Glazov:2020wf}. Significant differences between approaches appear for the effects related with the fine, spin-dependent, structure of the energy spectrum. This is because the trion, three-fermion complex, possesses a half-integer spin, while the Fermi polaron has an integer spin. Hence, these quasiparticles strongly differ, especially, if space- and time-reversal symmetries come into play. In our previous work~\cite{Iakovlev_2023} we have studied the role of anisotropic elastic deformations in the Fermi polaron and trion fine structure. We  have shown that in contrast to the trion, Fermi polaron states are split by anisotropic deformations into linearly polarized ones and the magnitude of the splitting is controlled by the anisotropic splitting of the neutral exciton and electron density vanishing in the limit of zero Fermi energy. 

The aim of this paper is to develop a theory of the energy spectrum fine structure of propagating Fermi polarons, i.e., for the quasiparticles having non-zero in-plane wavevector $\bm k$. We demonstrate that, like neutral excitons, the Fermi polarons are split into the longitudinal and transverse states which are linearly polarized along and perpendicular to the $\bm k$. We calculate the longitudinal-transverse splitting microscopically taking into account the long-range exchange interaction between the electron and hole or, in other words, the coupling of the Fermi polarons with the oscillating electromagnetic field produced by these optically active quasiparticles~\cite{BP_exch71,denisovmakarov,birpikus_eng}. We demonstrate that while the effect is similar to the longitudinal-transverse (LT) splitting of 2D excitons both in quantum wells~\cite{maialle93,goupalov98} and TMDC MLs~\cite{glazov2014exciton,Yu:2014fk-1,PhysRevB.89.205303,PSSB:PSSB201552211,prazdnichnykh2020control} it has substantial differences in molybdenum (Mo) and tungsten (W) based monolayers due to different arrangement of the spin subbands of conduction band.

The paper is organized as follows: after an introduction (Sec.~\ref{sec:intro}) the symmetry analysis of the trion and Fermi polaron fine structure is given in Sec.~\ref{sec:symm}. The microscopic model is presented in Sec.~\ref{sec:model} and the obtained results are discussed in Sec.~\ref{sec:disc}. The brief conclusion is presented in Sec.~\ref{sec:concl}.

\begin{figure*}[t]
    \centering
    \includegraphics[width=0.8\textwidth]{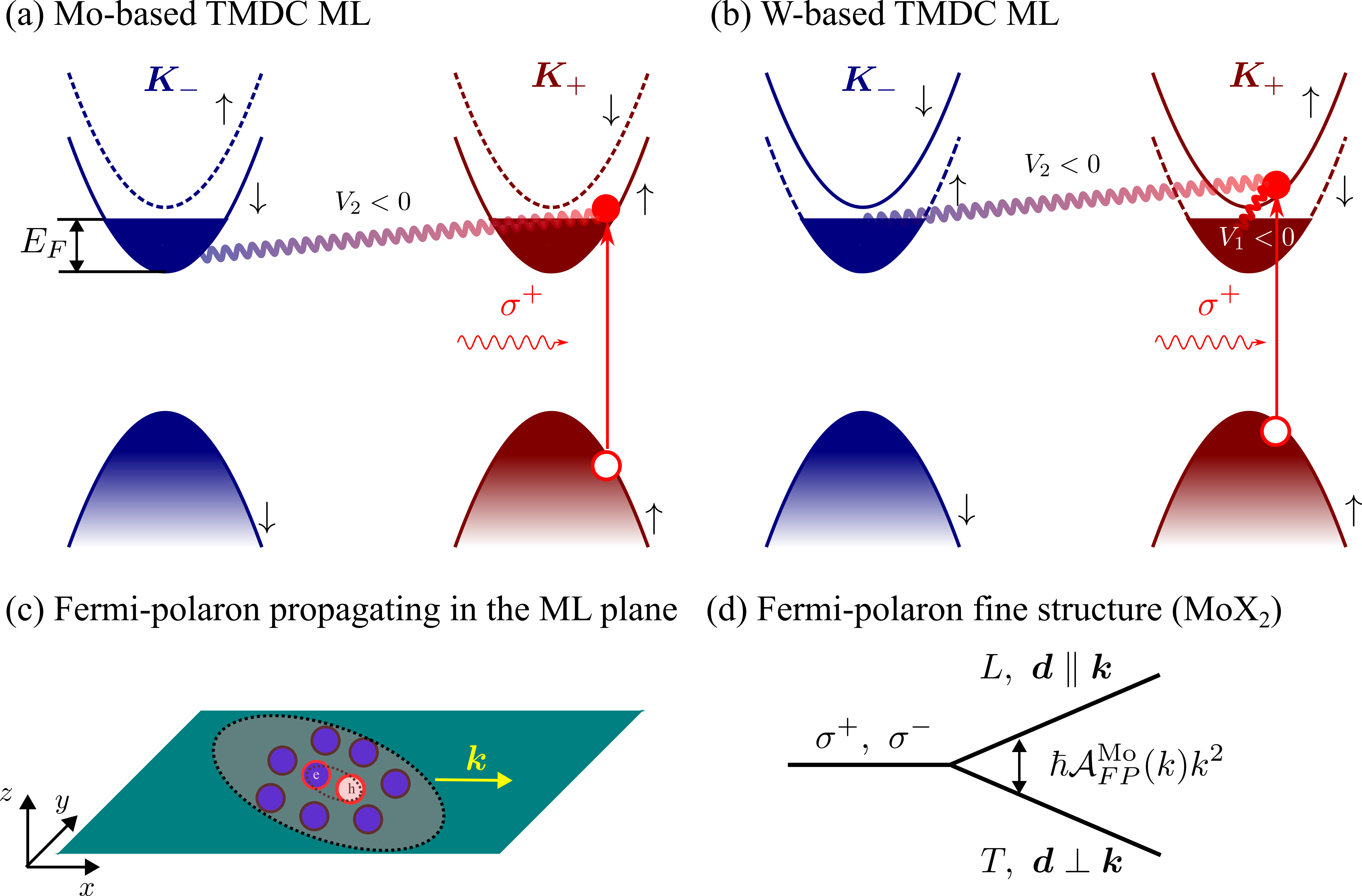}
    \caption{\textbf{Schematic illustration of Fermi polarons.} (a,b) Sketches of the band structure and relevant optical transitions for Mo-based (a) and W-based (b) TMDC MLs. The transition in $\bm K_+$ valley is induced by $\sigma^+$ polarized light and the transition in $\bm K_-$ is induced by the $\sigma^-$ polarized light (not shown). Wavy lines demonstrate intra- ($V_1$) and inter- ($V_2$) valley exciton-electron interaction, see Sec.~\ref{subsec:bare} for details. (c) Illustration of the Fermi polaron propagating in the ML plane; $\bm k$ is the Fermi polaron translational motion wavevector. (d) Schematics of the energy level splitting of propagating Fermi polaron in Mo-based ML, see Eqs.~\eqref{FP:FS:symm:Mo} and \eqref{FP:symm:split:Mo}. Here $\bm d$ is the microscopic dipole moment of the Fermi polaron whose direction controls the polarization of emission.}
    \label{fig:scheme}
\end{figure*}

\section{Symmetry analysis}\label{sec:symm}

Band structure in the vicinity of $\bm K_\pm$ points of the Brillouin zone and relevant optical transitions are schematically illustrated in Fig.~\ref{fig:scheme} where the panel (a) refers to the case of Mo-based TMDC MLs while the panel (b) refers to the W-based TMDC MLs. The key difference is the order of spin subbands in conduction band: for the Mo-based systems the topmost valence band and bottom conduction band have parallel spins ($+1/2$ in $\bm K_+$ valley and $-1/2$ in $\bm K_-$ valley, respectively) while in W-based systems the order of conduction bands is reversed~\cite{2053-1583-2-2-022001,Wang:2015aa,PhysRevB.93.121107}. As a result, despite apparent similarity of the band structure, the optical properties of such systems are qualitatively different. In Mo-based MLs the fundamental optical transition is spin allowed, while in W-based MLs it is forbidden. It results in different structure of the three-particle trion states in the case of the $n$-doped monolayers~\cite{Courtade:2017a,Iakovlev_2023}. Indeed, as it is well-known, the trion state for not too different effective masses of the electron and hole is bound provided that the envelope function of the relative motion of two electrons in the trion is symmetric with respect to their permutations~\cite{Courtade:2017a,PSSB:PSSB387}. Correspondingly, the Bloch function of two-identical particles should be antisymmetric. Thus, in Mo-based semiconductors two electrons forming the trions should be in different valleys, i.e., the trion is intervalley, Fig.~\ref{fig:scheme}(a). In W-based semiconductors two options are possible: either two electrons remain in the same valley but have opposite spins or the electrons stay in different valleys. Note that there is no Pauli restriction in the latter case because electrons are in different conduction bands. Correspondingly, inter- and intra-valley trions coexist in W-based TMDCs, Fig.~\ref{fig:scheme}(b). These trions have  somewhat different binding energies, see Refs.~\cite{Courtade:2017a,Zipfel:2020tq} and references therein for details. In the case of $p$-type doping both in MoX$_2$ and WX$_2$ MLs the trions are of the intervalley type because of the large spin-orbit splitting of the valence band. Note that hereafter we consider the situation of small or moderate doping where the Fermi energy of the charge carriers is smaller than the conduction (and valence) band spin splitting. We also focus on the $n$-type case since it is more general. As a result, only bottom spin sublevels of the conduction band are filled with electrons.

The trions are formed of three fermions, as a result, the spin of the trion is half-integer~\cite{Courtade:2017a,Iakovlev_2023}. It means that the time-reversal invariant perturbations such as, e.g., anisotropic elastic strain, cannot split the Kramers-degenerate (time-reversal related) states. If a trion propagates with a wavevector $\bm k$ in the monolayer plane, the broken space inversion symmetry of TMDC MLs (point group $D_{3h}$) allows for $k^3$ spin-dependent splitting stemming from the $k^3$ terms in the electron and hole dispersion~\cite{2053-1583-2-2-022001}. These contributions due to the spin-orbit interaction are disregarded in what follows.

The situation is different for Fermi polarons where the trion is correlated with the Fermi sea hole~\cite{Iakovlev_2023}. Since the spins of the electron in the Fermi sea and the hole in the Fermi sea are opposite, the Fermi polaron spin (or angular momentum) is the same as for the optically active exciton and equals to $m_z = \pm 1$ depending on the valley where the electron-hole pair is excited, Fig.~\ref{fig:scheme}(a,b). It is inherited from the helicity of photon absorbed by the ML semiconductor. As a result, the Fermi polaron inherits the fine structure from the constituting exciton. Following the method developed in Ref.~\cite{Iakovlev_2023} we obtain the following effective $2\times 2$ Hamiltonian describing the radiative doublet of Fermi polarons in Mo-based ML semiconductors 
\begin{subequations}
    \label{FP:FS:symm}
    \begin{equation}
    \label{FP:FS:symm:Mo}
        \hat{\mathcal H}_{\bm k}^{\rm Mo}  = \frac{\hbar\mathcal A_{FP}^{\rm Mo}(k)}{2} \left[(k_x^2 - k_y^2)\hat{\sigma}_x + 2 k_x k_y\hat{\sigma}_y  \right]{+ \hbar\mathcal{B}^{\rm Mo}_{FP}(k)\hat{I}}.
    \end{equation}
Here $x$ and $y$ are the in-plane coordinate axes, $k_x$ and $k_y$ are the corresponding Cartesian components of the Fermi polaron wavevector, $\hat \sigma_x$ and $\hat \sigma_y$ are the pseudospin Pauli matrices, and $\hat{I}$ is the unit $2\times 2$ matrix. Similarly to excitons and photons, the $x$-component of the pseudospin describes the linear polarization degree in the $(xy)$ axes frame and $y$-component describes the linear polarization degree in the $(x'y')$ frame rotated by $45^\circ$ with respect to the $(xy)$ axes; the $z$-component of the pseudospin gives the circular polarization degree~\cite{ivchenko05a,PSSB:PSSB201552211,PhysRevB.106.125303}. The function $\mathcal A_{FP}(k)$ describes the magnitude of the splitting and $\mathcal B_{FP}(k)$ describes the overall energy shift. Equation~\eqref{FP:FS:symm:Mo} has the same form as for the bright exciton radiative doublet~\cite{maialle93,glazov2014exciton}. Note that the Hamiltonian~\eqref{FP:FS:symm:Mo} acts in the basis of two intervalley Fermi polaron states with excitons in the $\bm K_+$ valley and in the $\bm K_-$ valley, respectively, see Fig.~\ref{fig:scheme}(a). To be specific, in this work we use the canonical basis for the representation of the circularly polarized states, hence, the $x$-polarized state is defined as $|x\rangle = (|\sigma^-\rangle - |\sigma^+\rangle)/\sqrt{2}$.

For W-based TMDC MLs the situation is somewhat different. There are, with account for polarization, four Fermi polaron states: two intervalley and two intravalley, Fig.~\ref{fig:scheme}(b). Each doublet is described by the Hamiltonian similar to Eq.~\eqref{FP:FS:symm:Mo}. In addition, symmetry permits the coupling between the inter- and intravalley polarons. The Hamiltonian describing four states takes the form of the $4\times 4$ matrix build from the blocks as
\begin{equation}
    \label{FP:FS:symm:W}
     \hat{\mathcal H}_{\bm k}^{\rm W}  = \begin{pmatrix}
         -\frac{\Delta}{2} \hat{I} +\hat{\mathcal H}_{\bm k}^{1} & {\hat{\mathcal H}}_{\bm k}^{12}\\
         {\hat{\mathcal H}}_{\bm k}^{12,\dag} & \frac{\Delta}{2} \hat{I}+ \hat{\mathcal H}_{\bm k}^{2}
     \end{pmatrix},
\end{equation}
where $\Delta$ is the splitting of the intra- and intervalley Fermi polarons (it weakly depends on $k$, but depends on the Fermi energy $E_F$, see Sec.~\ref{subsec:bare}), the $2\times 2$ matrices $\hat{\mathcal H}_{\bm k}^{1}$, $\hat{\mathcal H}_{\bm k}^{2}$, and $\hat{\mathcal H}_{\bm k}^{12}$ have the same form as Eq.~\eqref{FP:FS:symm:Mo} but with different prefactors denoted as $\mathcal A_{FP}^{(1)}$, $\mathcal A_{FP}^{(2)}$, $\mathcal A_{FP}^{(12)}$, $\mathcal B_{FP}^{(1)}$, $\mathcal B_{FP}^{(2)}$, and $\mathcal B_{FP}^{(12)}$ (the contribution with $\Delta$ is separated from the $\mathcal B_{FP}^{(1,2)}$ terms for convenience). Corresponding blocks describe the intra- and intervalley Fermi polaron states and their mixing. We assume that $\Delta>0$ and the lowest in energy doublet (block 1) corresponds to the Fermi polaron stemming from the intravalley (also known as singlet~\cite{Robert:2021wc}) trion. The block 2 is higher in energy and corresponds to the Fermi polaron with intervalley (triplet) trion. 
\end{subequations}

The Hamiltonian for the $p$-doped monolayers and for Fermi polarons in conventional semiconductor quantum wells has the same form as Eq.~\eqref{FP:FS:symm:Mo}.

It follows from Eqs.~\eqref{FP:FS:symm} that the Fermi polaron states are split, similarly to the exciton states, into linearly polarized combinations along and perpendicular to the wavevector $\bm k$, as illustrated in Fig.~\ref{fig:scheme}(d). The splitting is given by 
\begin{subequations}
    \label{FP:symm:split}
    \begin{equation}
        \label{FP:symm:split:Mo}
        \Delta_{FP}^{\rm Mo} = \hbar \mathcal  A_{FP}^{\rm Mo}(k) k^2,
    \end{equation}
for the molybdenum-based TMDC MLs, and
   \begin{equation}
        \label{FP:symm:split:W}
        \Delta_{FP}^{(1,2)} = \frac{\hbar \mathcal  A_{FP}^{(1)}(k) k^2}{2}+\frac{\hbar \mathcal   A_{FP}^{(2)}(k) k^2}{2}
        \mp  \left[\delta_1 - \delta_2\right],
   \end{equation}
   with 
   \begin{multline}
       \label{deltasmall12}
\delta_{1,2}^2 = \left(\hbar\mathcal{B}^{(12)}_{FP} \pm \frac{\hbar k^2\mathcal A_{FP}^{(12)}}{2}\right)^2 
\\+ \left(\frac{\Delta}{2} + \frac{\hbar\left[\mathcal{B}^{(2)}_{FP} - \mathcal{B}^{(1)}_{FP}\right]}{2}\pm \frac{\hbar k^2\left[\mathcal A_{FP}^{(2)} - \mathcal A_{FP}^{(1)}\right]}{4}\right)^2.
    \end{multline}
\end{subequations}

In the following section, we provide a microscopic model and calculate the parameters $\mathcal A_{FP}(\bm k)$ and $\mathcal B_{FP}(k)$ in phenomenological equations introduced above.

\section{Model}\label{sec:model}

Similarly to the calculation of the long-range exchange interaction effect on neutral exciton fine structure in bulk semiconductors~\cite{BP_exch71,denisovmakarov}, quantum wells~\cite{maialle93,goupalov98}, and two-dimensional TMDCs~\cite{glazov2014exciton,prazdnichnykh2020control} it is instructive to, first, find the energies and wavefunctions of Fermi polarons disregarding the interaction of these quasiparticles with induced electromagnetic field (Sec.~\ref{subsec:bare}) and then to account for the light-matter interaction using the perturbation theory (Sec.~\ref{subsec:LT}).

\subsection{Bare Fermi polarons}\label{subsec:bare}

We start with solving the quantum-mechanical problem of an exciton interacting with a Fermi sea of resident carriers. Following Ref.~\cite{Iakovlev_2023} we present the bare Fermi polaron (i.e., neglecting the coupling to the electromagnetic field) Hamiltonian as
\begin{multline}
\label{Ham:bare}
    \hat{\mathcal{H}}_0 = \sum_{\bm k}\varepsilon_{\bm k}^X\left(\hat{R}_{\bm k}^\dagger\hat{R}_{\bm k} + \hat{L}_{\bm k}^\dagger\hat{L}_{\bm k}\right) \\ + \sum_{\bm k}\varepsilon_{\bm k}\left(\hat{r}^\dagger_{\bm k}\hat{r}_{\bm k} + \hat{l}^\dagger_{\bm k}\hat{l}_{\bm k}\right) \\ + V_1\sum_{\bm k, \bm k', \bm p, \bm p'}\delta_{\bm k + \bm p, \bm k' + \bm p'}\left(\hat{R}^\dagger_{\bm k'}\hat{r}^\dagger_{\bm p'}\hat{R}_{\bm k}\hat{r}_{\bm p} + \hat{L}^\dagger_{\bm k'}\hat{l}^\dagger_{\bm p'}\hat{L}_{\bm k}\hat{l}_{\bm p}\right) \\ + V_2\sum_{\bm k, \bm k', \bm p, \bm p'}\delta_{\bm k + \bm p, \bm k' + \bm p'}\left(\hat{R}^\dagger_{\bm k'}\hat{l}^\dagger_{\bm p'}\hat{R}_{\bm k}\hat{l}_{\bm p} + \hat{L}^\dagger_{\bm k'}\hat{r}^\dagger_{\bm p'}\hat{L}_{\bm k}\hat{r}_{\bm p}\right).
\end{multline}
Here $\hat{R}_{\bm k}^\dag$, $\hat{R}_{\bm k}$ and $\hat{L}_{\bm k}^\dag$, $\hat{L}_{\bm k}$ are the creation and annihilation operators of excitons active, respectively, in the right ($R$) and left ($L$) circular polarizations, i.e., corresponding to the optical transitions from the valence to conduction band in the $\bm K_+$ and $\bm K_-$ valleys, $\hat{{r}}_{\bm k}^\dag$, $\hat{{r}}_{\bm k}$ and $\hat{{l}}_{\bm k}^\dag$, $\hat{{l}}_{\bm k}$ are the same operators for electrons in the $\bm K_\pm$ valleys, $\varepsilon^X_{\bm k} = \h^2k^2 / (2M_X)$ and $\varepsilon_{\bm k} = \h^2k^2 / (2M_e)$ are the kinetic energies of excitons and electrons, respectively, with their translational masses being $M_X$ and $M_e$. The parameters $V_1$ and $V_2$ describe intra-valley and inter-valley interaction, respectively. They are related to the trion binding energy, see below. Kronecker $\delta$-symbol describes the momentum conservation law at the electron-exciton scattering. The Hamiltonian~\eqref{Ham:bare} is written for the W-based system, where both inter- and intravalley trions are bound and $V_{1,2} < 0$. In the case of Mo-based system (or conventional quantum well), formally, the parameter $V_1 > 0$ is positive due to exchange interaction and can be disregarded (set to zero) in further derivations.  In this case, intravalley trions and corresponding Fermi polarons are absent~\cite{Iakovlev_2023}. We note that the Hamiltonian~\eqref{Ham:bare} includes the static Coulomb interaction resulting in formation of mechanical excitons and the intraband exchange interaction crucial for formation of \emph{symmetric} trions (with the envelope function being symmetric with respect to permutation of electrons~\cite{Courtade:2017a}). Equation~\eqref{Ham:bare} disregards the interband, long-range exchange interaction related to virtual annihilation and creation of electron-hole pairs, which is analyzed below in Sec.~\ref{subsec:LT}.

Neglecting the valley mixing due to the long-range exchange interaction, the Fermi polaron states active in $\sigma^+$ and $\sigma^-$ circular polarizations are independent. The $\sigma^+$ polarized state reads~\cite{suris:correlation,PhysRevA.74.063628,Sidler:2016aa,Iakovlev_2023}
\begin{multline}
\label{Psi:k:+}
    \left|\Psi_{\bm k}^{\sigma^+}\right\rangle = \varphi_{\bm k}R_{\bm k}^\dagger|FS\rangle \\ + \sum_{\bm p, \bm q}F_{\bm k}^{RR}(\bm p, \bm q)\hat{R}_{\bm k - \bm p + \bm q}^\dagger \hat{r}_{\bm p}^\dagger \hat{r}_{\bm q}|FS\rangle \\ + \sum_{\bm p, \bm q}F_{\bm k}^{RL}(\bm p, \bm q)\hat{R}_{\bm k - \bm p + \bm q}^\dagger \hat{l}_{\bm p}^\dagger \hat{l}_{\bm q}|FS\rangle,
\end{multline}
where $\bm k$ is the Fermi polaron momentum, $\varphi_{\bm k}$ describes the excitonic part of the Fermi polaron (with $|\varphi_{\bm k}|^2 \ll 1$ for the attractive~-- trion-like~-- Fermi polaron), $F^{RR}_{\bm k}(\bm p, \bm q)$ and $F^{RL}_{\bm k}(\bm p, \bm q)$ are the coefficients describing the intra- and inter-valley admixtures of electron-hole pair excitations to the right circularly polarized excitons. In Eq.~\eqref{Psi:k:+} $|FS\rangle$ is the state of unperturbed Fermi sea with bottom spin subbands of conduction bands filled with electrons up to the Fermi energy $E_F$, see Fig.~\ref{fig:scheme}. The Fermi sea hole should be below the Fermi level, and the electron excited from the Fermi sea should be above the Fermi level. Hence, hereafter we assume summation over $\bm p, \bm p', \bm q, \bm q'$ with $p, p' \geq k_F \geq q, q'$, where $k_F = \sqrt{2M_eE_F}/\h$ is the Fermi wavevector. The wave function for left circularly polarized Fermi polaron, $\left|\Psi_{\bm k}^{\sigma^-}\right\rangle$, can be obtained from Eq.~\eqref{Psi:k:+} by the replacements $R \leftrightarrow L$ and $r \leftrightarrow l$. In the absence of the long-range exchange interaction the wavefunctions $\left|\Psi_{\bm k}^{\sigma^+}\right\rangle$ and $\left|\Psi_{\bm k}^{\sigma^-}\right\rangle$ have exactly the same energies, the exchange interaction will mix and split these states, see Sec.~\ref{subsec:LT}. 

Following Ref.~\cite{Iakovlev_2023} to find the Fermi polaron dispersion we substitute $\left|\Psi_{\bm k}^{\sigma^+}\right\rangle$ from Eq.~\eqref{Psi:k:+} to the Schr\"odinger equation 
\begin{equation}
    \hat{\mathcal H}_0 \left|\Psi_{\bm k}^{\sigma^+}\right\rangle = E_{\bm k} \left|\Psi_{\bm k}^{\sigma^+}\right\rangle,
\end{equation} 
and apply variational principle treating the coefficients $\varphi_{\bm k}$, $F_{\bm k}^{RR}$, and $F_{\bm k}^{RL}$ are variational parameters. It gives the equation for the Fermi polaron energy
\begin{equation}
    \sum_{\bm q}\frac{V_1}{1 - V_1S_{\bm k}(\bm q)} + \sum_{\bm q}\frac{V_2}{1 - V_2S_{\bm k}(\bm q)} + \eps^X_{\bm k} = E_{\bm k},
    \label{eq:eigenvalue}
\end{equation}
where we introduced the function $S_{\bm k}(\bm q)$ as the sum of the Green's function of non-interacting electron-exciton pair $\zeta_{\bm k}(\bm p, \bm q)$
\begin{subequations}
\begin{equation}
    S_{\bm k}(\bm q) = \sum_{p \geq k_F}\zeta_{\bm k}(\bm p, \bm q),
\end{equation}
\begin{equation}
    \zeta_{\bm k}(\bm p, \bm q) = \frac1{E_{\bm k} - \eps^X_{\bm k - \bm p + \bm q} - \eps_{\bm p} + \eps_{\bm q}}.
\end{equation}
The zero energy for the quasiparcticles, $E_{\bm k} = 0$, corresponds to the bare exciton energy at $E_F = 0$.

The first two terms in Eq.~\eqref{eq:eigenvalue} have a physical meaning of the exciton self-energy related to the interaction with the Fermi sea~\cite{suris:correlation,Glazov:2020wf}
\begin{equation}
    \label{exciton:self}
    \Sigma_{\bm k}^{FS}(E_{\bm k}) = \sum_{\bm q}\frac{V_1}{1 - V_1S_{\bm k}(\bm q)} + \sum_{\bm q}\frac{V_2}{1 - V_2S_{\bm k}(\bm q)}.
\end{equation}
\end{subequations}
All wave function parameters can be expressed through the exciton amplitude $\varphi_{\bm k}$
\begin{equation}
    F^{RR}_{\bm k}(\bm p, \bm q) = \frac{V_1\zeta_{\bm k}(\bm p, \bm q)\varphi_{\bm k}}{1 - V_1S_{\bm k}(\bm q)},~~ F^{RL}_{\bm k}(\bm p, \bm q) = \frac{V_2\zeta_{\bm k}(\bm p, \bm q)\varphi_{\bm k}}{1 - V_2S_{\bm k}(\bm q)},
\end{equation}
which is determined by the state normalization condition $\langle \Psi_{\bm k}|\Psi_{\bm k}\rangle = 1$. Since we disregard $k^3$ and higher order in $\bm k$ contributions to the bare electron and exciton energies related to the absence of the inversion center of the TMDC monolayer, our model is essentially centrosymmetric. As a result, both intravalley and intervalley parts of the wave function are equal for the right and left circularly polarized Fermi polarons, \begin{equation}
    \label{rel:F}
    F^{RR}_{\bm k}(\bm p, \bm q) = F^{LL}_{\bm k}(\bm p, \bm q), \quad F^{RL}_{\bm k}(\bm p, \bm q) = F^{LR}_{\bm k}(\bm p, \bm q),
\end{equation}
which will be used further.

We introduce the trion binding energies $E_{T1,2}$ for the intra- and intervalley trions in W-based TMDC monolayers as~\cite{Iakovlev_2023}
\begin{equation}
    \label{ET12}
    E_{T1,2} = E_X \exp{\left(\frac{1}{\mathcal D V_{1,2}}\right)},
\end{equation}
where $E_X$ is the exciton binding energy and $\mathcal D $  the exciton-electron reduced density of states. In this model description it is assumed that $|\mathcal D V_{1,2}| \ll 1$ resulting in $E_{T1,2} \ll E_X$. For Mo-based TMDCs, the intravalley trion is absent, and we denote the intervalley trion binding energy as $E_T \equiv E_{T2}$ in this case. For the tungsten-based TMDCs, the intra- and inter-valley trion binding energies $E_{T1}, E_{T2}$ are relatively close to each other, such that $\Delta = |E_{T1} - E_{T2}| \ll E_T = (E_{T1} + E_{T2}) / 2$. Generally, this parameter $\Delta$ differs from the splitting of Fermi polarons at $\bm k=0$ introduced in Hamiltonian~\eqref{FP:FS:symm:W}, see below.

We are interested in the attractive (stemming from the trion) Fermi polaron dispersion. Let us start the analysis of the bare Fermi polaron dispersion for Mo-based monolayers. In the considered limit of high binding energy, $E_F, \hbar^2 k^2/2M_e \ll E_T$, the self-energy $\Sigma_{\bm k}^{FS}$, Eq.~\eqref{exciton:self}, takes form
\begin{equation}
        \label{Self}
        \Sigma_{\bm k}^{FS} = \left(\frac{M_T}{M_X}\right)^2E_T \ln{\left(1+\frac{\mathcal E_{1,\bm k}}{2\mathcal E_{2,\bm k}}\right)},
\end{equation}
where
\begin{subequations}
\begin{align}
    &\mathcal E_{1,\bm k} = \frac{M_X}{M_T}E_F \\
    &- \sqrt{\left(\Delta E + \frac{M_X}{M_T}E_F\right)^2 - 2\frac{M_e}{M_T}\frac{\h^2k^2}{M_T}E_F -\rm i0} - \Delta E, \nonumber\\
    &\mathcal E_{2,\bm k}  = \Delta E - \frac{M_e}{M_X}\frac{\hbar^2k^2}{2M_T},
\end{align}
\end{subequations}
where we introduced the trion effective mass $M_T = M_X + M_e$ and $\Delta E = E_{\bm k} - \left(-E_T + \frac{M_T}{M_X}E_F + \frac{\h^2k^2}{2M_T}\right)$. In Eq.~\eqref{Self} the square root is taken in such a way, that its imaginary part is negative or zero.
%
%
Solving Eq.~\eqref{eq:eigenvalue}, we get the attractive Fermi polaron dispersion for $k \ll k_F$ in the form
\begin{subequations}
    \label{FP:disper:bare}
\begin{equation}
    E_{\bm k} = -E_0 + \left\{\frac{M_T}{M_X} - \frac{M_X/M_T}{1 - \exp\left[-\left(\frac{M_X}{M_T}\right)^2\right]}\right\}E_F + \frac{\h^2k^2}{2M_{FP}}.
\end{equation}
Here, for Mo-based systems $E_0 = E_T$, the trion binding energy, and the attractive Fermi polaron effective mass $M_{FP}$ is given by
\begin{equation}
\label{MFP}
    \frac1{M_{FP}} = \frac1{M_T} - \frac{M_e}{M_XM_T}\left\{\exp\left[\left(\frac{M_X}{M_T}\right)^2\right] - 1\right\}.
\end{equation}
\end{subequations}
Interestingly, it is larger than the trion mass; for example, $M_{FP} \approx 4.17M_e$ in the case of equal masses of an electron and a hole, $M_X = 2M_e$.  

\begin{figure*}[t]
    \centering
    \includegraphics[width=0.8\textwidth]{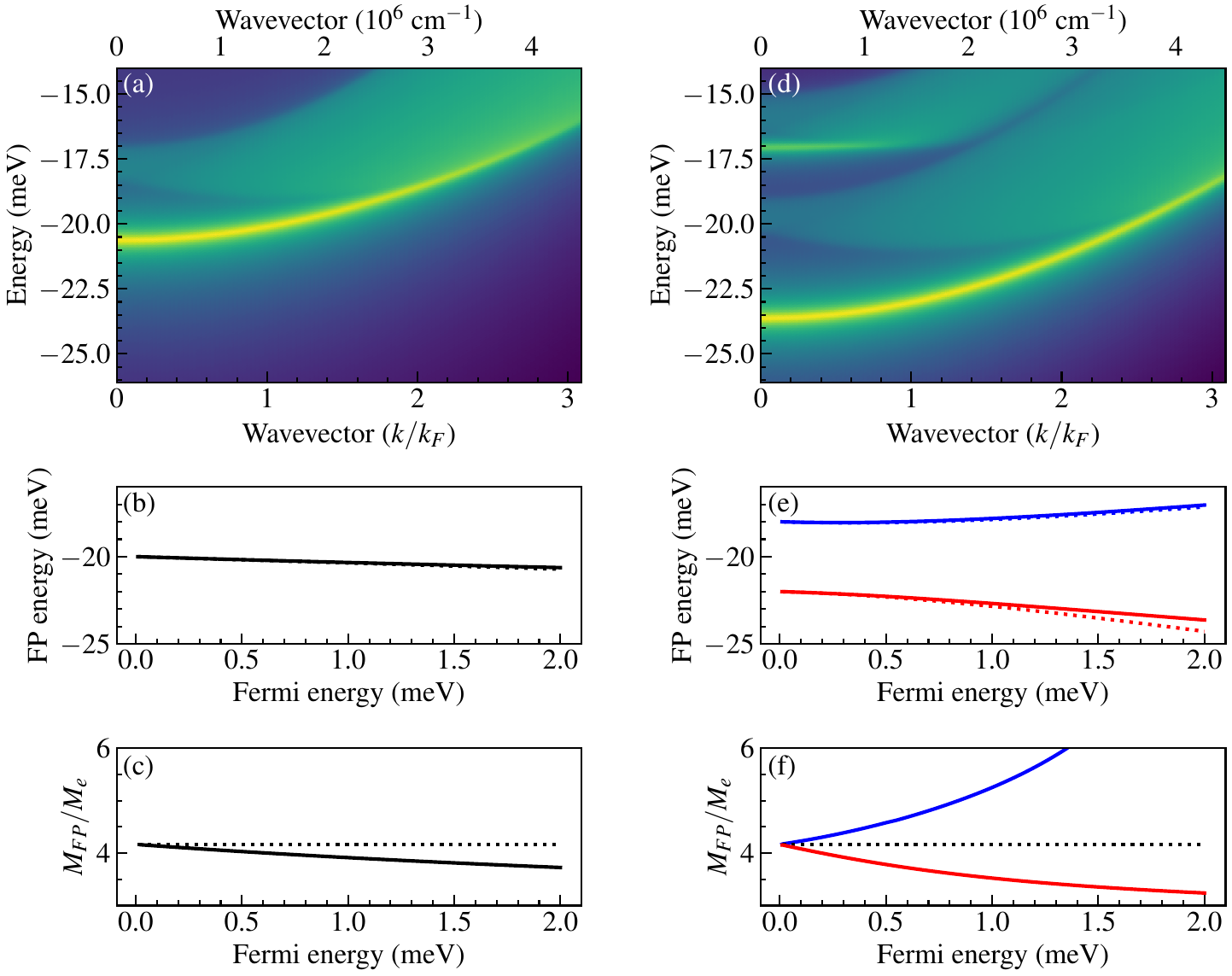}
    \caption{\textbf{Dispersion of Fermi polarons in absence of the long-range electron-hole exchange interaction.} (a-c) Mo-based monolayer, (d-f) W-based monolayer. Panels (a, d) show the density plot of $-\Im G_E(\bm k)$, Eq.~\eqref{G:E:FP}, in the logarithmic color scale. Panels (b, e) show the attractive polarons energy at $\bm k = 0$: solid lines are the results of numerical calculation and dotted lines are the analytical asymptotics, Eqs. \eqref{FP:disper:bare}. Panels (c, f) show the Fermi polaron mass found by the fitting of the dispersion obtained numerically from the poles of $G_E(\bm k)$ by the parabolic law (solid lines) and analytical Eq.~\eqref{MFP} (dotted lines). The parameters used in calculations are: $M_e = M_h = 0.4m_0$, $E_T = 20$meV, $E_F = 2$meV, $\gamma = 0.1$meV, $\Delta = 4$meV.}
    \label{fig:disper}
\end{figure*}

Calculations show that Eqs.~\eqref{FP:disper:bare} hold also for W-based TMDC monolayers. They describe the dispersion of both intra- and interlayer Fermi polarons with $E_0 = E_T \pm \sqrt{\Delta^2/4 + (\alpha E_F)^2}$ [strictly speaking, the combination $2\sqrt{\Delta^2/4 + (\alpha E_F)^2}$ enters Eq.~\eqref{FP:FS:symm:W} as $\Delta$]. Here the parameter
\begin{equation}
    \alpha = \frac{(M_X/M_T)^3}{4\sinh^2\left[\frac{1}{2}\left(\frac{M_X}{M_T}\right)^2\right]},
\end{equation}
and the exciton contribution to the attractive Fermi polaron that determines its oscillator strength is (cf. Ref.~\cite{Iakovlev_2023})
\begin{equation}
\label{phi:weight}
    \varphi_{\bm k} = \sqrt{\frac{\alpha E_F - \exp\left[\left(\frac{M_X}{M_T}\right)^2\right]\frac{M_eM_X}{M_T^2}\frac{\h^2k^2}{2M_T}}{E_T}}.
\end{equation}
One can notice that $\varphi_{\bm k}$ is well-defined only for sufficiently small wavevectors $k \lesssim k_F$, otherwise the expression under the square root is negative and the Fermi polaron enters the continuum of the uncorrelated trion-Fermi sea hole excitations~\cite{suris:correlation}.

Figure~\ref{fig:disper} shows the dispersion of bare attractive Fermi polarons. Left and right columns correspond to the Mo- and W-based monolayers, respectively. Panels (a) and (d) show the spectral density of Fermi polarons determined from the imaginary part of the exciton Greens function~\cite{Glazov:2020wf}
\begin{equation}
    \label{G:E:FP}
    G_E(\bm k) = \frac{1}{E - \varepsilon_{\bm k}^X - \Sigma_{\bm k}^{FS}(E+i\gamma) + \mathrm i \gamma},
\end{equation}
where $\gamma$ is the phenomenological damping. At $\bm k=0$ the Fermi polaron spectrum is similar to previously studied~\cite{suris:correlation,Glazov:2020wf,PhysRevB.95.035417,Iakovlev_2023}. For Mo-based TMDCs it contains the attractive polaron bound state and the continuum of uncorrelated trion-hole pairs in the Fermi sea. For W-based TMDCs two bound states related to the inter- and intravalley trions are seen and there are two corresponding continua. The Fermi polaron energies at $\bm k=0$ as functions of the electron Fermi energy are shown in Fig.~\ref{fig:disper}(b,e). At $k>0$ the Fermi polarons shift to the higher energy demonstrating parabolic dispersion at small momenta, Eqs.~\eqref{FP:disper:bare},  that becomes non-parabolic at higher $k$ and merges with the continua of uncorrelated states. We stress that the states containing excitons in $\bm K_+$ and $\bm K_-$ valleys are decoupled in this approximation. The mixing of the states appears as a result of the light-matter interaction; it is described in the next subsection.

\subsection{Fine structure}\label{subsec:LT}

Microscopically, the fine structure of neutral excitons and Fermi polarons described by phenomenological Eqs.~\eqref{FP:FS:symm} appears with account for the exciton interaction with induced electromagnetic field. Quantum-mechanically, it can be described as a result of virtual annihilation and creation of the electron-hole pair with photon emission and absorption. In the macroscopic electrodynamics approach, it corresponds to inclusion of polariton effects, while the Fermi polarons found in Sec.~\ref{subsec:bare} correspond to ``mechanical excitons''. Here, as we demonstrate below, it is convenient to simultaneously take into account the coupling for polarons with both longitudinal and transverse components of the electromagnetic field~\cite{agr_ginz}.

The exciton self-energy related to the virtual annihilation and creation process can be written as~\cite{glazov2014exciton,prazdnichnykh2020control}
\begin{multline}
    \label{exciton-self:fin}
    \Sigma_{\alpha\beta}(\omega,\bm k) = - \hbar \Gamma_0 \\
    \times \left(\delta_{\alpha\beta} - \frac{k_\alpha k_\beta}{(\omega/c)^2} \right) \frac{(\omega/c)}{\sqrt{k^2-(\omega/c)^2 - \mathrm i0}},
\end{multline}
where $\alpha,\beta=x,y$ are the Cartesian indices corresponding to the orientation of the microscopic dipole moment of the exciton in the monolayer plane, $\omega$ is the frequency of the electromagnetic field corresponding to the quasiparticle energy $\omega = E/\hbar$, where $E \approx E_g - E_X$, and $\Gamma_0$ is the bright exciton radiative decay rate. Up to a constant prefactor $\hbar\Gamma_0$, the self-energy~\eqref{exciton-self:fin} is the Green's function of electromagnetic field (in the gauge where the scalar potential is zero) at the $z=0$ corresponding to the position of the monolayer occupying the $(xy)$ plane, Fig.~\ref{fig:scheme}(c). We assume that the monolayer is in the vacuum, the allowance for the dielectric surrounding can be carried out in a standard way~\cite{prazdnichnykh2020control,PhysRevLett.123.067401,ren2023control}. The self-energy~\eqref{exciton-self:fin} takes into account coupling with all modes of electromagnetic field. Interaction with longitudinal field corresponds to the limit $k\equiv \sqrt{k_x^2+k_y^2}\gg \omega/c$ where 
\begin{equation}
\label{sigma:long}
\Sigma_{\alpha\beta} = \hbar\Gamma_0 \frac{k_\alpha k_\beta}{k(\omega/c)},  
\end{equation}
i.e., where the retardation related to the finite speed of light can be neglected.

As expected, for bright exciton doublet the self-energy corresponds to the effective $2\times 2$ Hamiltonian describing the radiative doublet in the form of Eq.~\eqref{FP:FS:symm:Mo} with excitonic function 
\begin{equation}
    \label{AXk}
    \mathcal A_X(k) = -\Gamma_0(\omega k/c)^{-1},
\end{equation} 
for the states outside the light cone in agreement with previous works~\cite{glazov2014exciton,prazdnichnykh2020control,PhysRevB.106.125303} (the sign $-$ in Eq.~\eqref{AXk} is related to the choice of the canonical basis, see Sec.~\ref{sec:symm}). The overall exciton energy shift owing to the interaction with electromagnetic field (analogue of the Lamb shift~\cite{ren2023control}) is $\hbar\mathcal{B}_X(k) = (kc/\omega)\hbar\Gamma_0 / 2$. A combination of the longitudinal-transverse  splitting, $\propto \mathcal A$, and the shift, $\propto \mathcal B$, gives the well-known picture of the transverse exciton being practically intact by the long-range exchange interaction and of the longitudinal one being pushed to higher energies~\cite{BP_exch71,PhysRevB.41.7536}.
For the states within the light cone Eq.~\eqref{exciton-self:fin} properly describes the radiative decay~\cite{glazov2014exciton,PhysRevLett.123.067401}; particularly, at $k=0$ we have $\Im\Sigma_{\alpha\beta}(\omega, 0) = -\mathrm i \delta_{\alpha\beta} \times {\hbar}\Gamma_0$. 

It is instructive to recast Eq.~\eqref{exciton-self:fin} in the second quantization representation via the creation and annihilation operators of the circularly polarized excitons introduced in Eq.~\eqref{Ham:bare}. The corresponding perturbation Hamiltonian takes the form
\begin{multline}
\label{Ham:X:pert}
    \hat{\mathcal H}_{X}  = \hbar\Gamma_0\sum_{\bm k}\frac{\omega / c}{\sqrt{k^2 - (\omega / c)^2 - \rm i0}} \\ 
    \times \biggl[\left(\frac{k^2}{2(\omega / c)^2} - 1\right)\left(\hat{L}^\dag_{\bm k}\hat{L}_{\bm k} + \hat{R}^\dag_{\bm k}\hat{R}_{\bm k}\right) \\ - \frac{(k_x - \mathrm i k_y)^2}{2(\omega / c)^2}\hat{L}^\dag_{\bm k}\hat{R}_{\bm k} - \frac{(k_x + \mathrm ik_y)^2}{2(\omega / c)^2}\hat{R}^\dag_{\bm k}\hat{L}_{\bm k}\biggr].
\end{multline}
Hence, to find the fine structure of Fermi polarons microscopically, we evaluate the matrix elements of Eq.~\eqref{Ham:X:pert} using the bare Fermi polaron wavefunctions $\left|\Psi_{\bm k}^{\sigma^+}\right\rangle$ and $\left|\Psi_{\bm k}^{\sigma^-}\right\rangle$, Eq.~\eqref{Psi:k:+}.

For Mo-based monolayer and in the case of $p$-doping we immediately obtain
\begin{subequations}
    \label{results:Mo}
\begin{equation}
    \label{splitting:Mo}
    \mathcal A_{FP}^{\rm Mo}(k) = |\varphi_k|^2 \mathcal A_X(k) \approx -|\varphi_k|^2\frac{\Gamma_0}{(\omega k/c)},
\end{equation}
\begin{equation}
\label{shift:Mo}
\mathcal B_{FP}^{\rm Mo}(k) =  |\varphi_k|^2 \mathcal B_X(k) \approx  |\varphi_k|^2 k \frac{ \Gamma_0}{2(\omega/c)},
\end{equation}
\end{subequations}
where the exciton contribution to the attractive Fermi polaron is given by Eq.~\eqref{phi:weight}. It can be roughly estimated as the ratio of the Fermi energy to the trion binding energy, $E_F/E_T$. Hence, according to Eq.~\eqref{FP:symm:split:Mo} the Fermi polaron longitudinal-transverse splitting is linear in the wavevector $\bm k$ (for the states with $k\gg \omega/c$ but $k \ll k_F$) and is proportional to the exciton splitting with a coefficient $E_F/E_T \ll 1$:
\begin{equation}
    \label{results:Mo:1}
    \Delta_{FP}^{\rm Mo} \approx  - \frac{\alpha E_F}{E_T}k\frac{\hbar \Gamma_0}{(\omega/c)}.
\end{equation}
Equation~\eqref{splitting:Mo} is similar to that describing the strain-induced splitting in Mo-based TMDC MLs~\cite{Iakovlev_2023}: The attractive Fermi polaron ``inherits'' the splitting from the exciton, hence, the splitting is proportional to the exciton fraction in the Fermi polaron, $|\varphi_k|^2$. The `$-$' sign means, that the longitudinal Fermi polaron has higher energy than the transversal one.

For W-based monolayer the situation is more involved because of the mixing between the intra- and intervalley trions in Fermi polarons. In this case, a non-zero contribution to $\mathcal A_{FP}^{\rm W}$ results from the virtual recombination and annihilation of exciton in the presence of the electron-hole pair [i.e., second and third terms in Eq.~\eqref{Psi:k:+}. Calculation shows that
\begin{subequations}
\label{results:W}
\begin{multline}
    \label{splitting:W}
    \mathcal A_{FP}^{(1,2)}{(k)} = |\varphi_k|^2 \mathcal A_X{(k)} 
    \\ + \sum_{\bm p, \bm q}\sum_{\sigma\in \{R, L\}}F_{\bm k}^{(1,2),R\sigma}(\bm p, \bm q)F_{\bm k}^{(1,2),L\sigma}(\bm p, \bm q)\\\times \mathcal{A}_X(|\bm k - \bm p + \bm q|)\left(1 - \frac{p_x - q_x}{k_x}\right)^2,
\end{multline}
\begin{multline}
    \mathcal{B}^{(1,2)}_{FP}(k) = |\varphi_k|^2\mathcal{B}_X(k) \\ + \sum_{\bm p, \bm q}\sum_{\sigma\in \{R,L\}}\left[F^{(1,2),R\sigma}_{\bm k}(\bm p, \bm q)\right]^2
    \\
    \times \mathcal{B}_x(|\bm k - \bm p + \bm q|).
\end{multline}
\begin{multline}
    \mathcal A_{FP}^{(12)}(k) = |\varphi_k|^2 \mathcal A_X(k) 
    \\ + \sum_{\bm p, \bm q}\sum_{\sigma\in \{R, L\}}F_{\bm k}^{(1),R\sigma}(\bm p, \bm q)F_{\bm k}^{(2),L\sigma}(\bm p, \bm q)\\ \times \mathcal{A}_X(|\bm k - \bm p + \bm q|)\left(1 - \frac{p_x - q_x}{k_x}\right)^2,
\end{multline}
\begin{multline}
    \mathcal{B}^{(12)}_{FP}(k) = |\varphi_k|^2\mathcal{B}_X(k) \\ + \sum_{\bm p, \bm q}\sum_{\sigma\in \{R,L\}}F^{(1),R\sigma}_{\bm k}(\bm p, \bm q)F^{(2),R\sigma}_{\bm k}(\bm p, \bm q)\\ \times \mathcal{B}_x(|\bm k - \bm p + \bm q|).
\end{multline}
\end{subequations}
Equations~\eqref{results:W} present the general expressions for the fine structure of Fermi polarons (Suris tetrons) in W-based monolayers. These expressions can be strongly simplified in the leading order in $E_F/\Delta$ and for small $k$, but, as above, larger than $\omega/c$: 
\begin{subequations}
\label{results:W:1}
\begin{equation}
    {\mathcal{A}^{(1,2)}_{FP}(k)} = -\frac{\alpha E_F}{E_T}\frac{\Gamma_0}{(\omega k/c)}\mp\frac{3\pi}{8}\frac{\alpha E_F}{\Delta}\sqrt{\frac{\frac{\hbar^2}{2\mu_{eX}}}{E_T}}\frac{c}{\omega}\Gamma_0,
\end{equation}
\begin{equation}
\label{A:12:contrib}
    \mathcal{A}^{(12)}_{FP}(k) = -\frac{\alpha E_F}{E_T}\frac{\Gamma_0}{(\omega k/c)}-\frac{3\pi}{16}\sqrt{\frac{\frac{\hbar^2}{2\mu_{eX}}}{E_T}}\frac{c}{\omega}\Gamma_0,
\end{equation}
\begin{equation}
    \mathcal{B}^{(1,2)}_{FP}(k) =  \left(\frac{\alpha E_F}{2E_T} + \frac{\pi}{16}\sqrt{\frac{\frac{\hbar^2k^2}{2\mu_{eX}}}{E_T}}\right)\frac{kc}{\omega}\Gamma_0 + \frac{\pi}{4}\sqrt{\frac{E_T}{\frac{\hbar^2(\omega/c)^2}{2\mu_{eX}}}}\Gamma_0,
\end{equation}
\begin{equation}
    {\mathcal{B}^{(12)}_{FP}(k) = \frac{\alpha E_F}{2E_T}\frac{kc}{\omega}\Gamma_0.}
\end{equation}
\end{subequations}
Here $\mu_{eX} = M_eM_X / M_T$ is electron-exciton reduced mass. First terms in Eqs.~\eqref{results:W:1} [and the only contribution to $\mathcal{B}^{(12)}_{FP}(k)$] result from the excitonic contribution to the Fermi polaron wavefunction. These contributions are similar to those in Mo-based TMDC MLs and contain a small factor $E_F/E_T$. Other contributions do not have this small parameter $E_F/E_T$, these contributions are specific for the W-based MLs. Importantly, the momentum dependence of these contributions is different, as addressed in more detail in Sec.~\ref{sec:disc}.  Interestingly, in the polarization-independent mixing parameter $\mathcal{B}^{(12)}_{FP}(k)$ the terms related to the admixture of the inter- and intravalley trions cancel each other and only the excitonic contribution remains.

It is worth noting that the parameter $\mathcal A_{FP}^{(12)}$ in Eq.~\eqref{A:12:contrib} contains contribution which does not depend on the Fermi energy. Physically, it is related to the mixing of the inter- and intravalley trions by the long-range exchange interaction. In such a case, an exciton changes the valley, but the resident electron does not. As it is readily seen from Eq.~\eqref{FP:symm:split} this contribution itself (i.e., where other terms are absent) does not result in a longitudinal-transverse splitting of Fermi polarons. It effectively renormalizes the splitting between inter- and intravalley trions $\Delta$. In our calculations, we also exclude the overall $k$- and $E_F$-independent energy shift of Fermi polarons due to the last term in $\mathcal B^{(1,2)}_{FP}(k)$, as it effectively renormalizes the trions binding energies and can be included in $E_{T1,2}$ together with the short-range exchange interaction part~\cite{Courtade:2017a}.

Note that Eqs.~\eqref{results:W} and \eqref{results:W:1} are derived neglecting retardation, i.e., assuming that all relevant wavevectors $k,p,q$ in summations exceed the light wavevector $\omega/c$. Inclusion of retardation allows one to describe also the radiative decay of the Fermi polarons including the ``recoil'' effect~\cite{PhysRevB.105.075311}\, but leads to minor variations of the fine structure splittings neglected here because both exciton and trion `Bohr' radii as well as the Fermi wavelength are much smaller than the  wavelength of light $c/\omega$.

Neglecting the terms $\propto E_F/E_T$ we simplify the expressions for the splittings of the Fermi polarons in tungsten-based TMDC MLs to 
\begin{equation}
\label{Delta:FP:W}
   \Delta_{FP}^{(1)}  = -  \Delta_{FP}^{(2)} = \delta_{2} - \delta_{1},
\end{equation}
where, in agreement with Eq.~\eqref{deltasmall12},
\begin{equation}
\label{deltasmall12:1}
    \delta_{1,2} = \left[\left(\frac{\Delta}{2} \mp \frac{\hbar \mathcal{A}^{(1)}_{FP}(k)k^2}{2}\right)^2 \right.
    + \left. \left(\frac{\hbar \mathcal A_{FP}^{(12)}(k)k^2}{2}\right)^2\right]^{1/2},
\end{equation}
and $\delta_1>\delta_2$.
In this limit, like in the Mo-based system, the situation is similar to that in the presence of anisotropic strain. The leading contributions to the Fermi polaron fine structure result from the mixing of the intra- and intervalley contributions in the Fermi polarons~\cite{Iakovlev_2023}. 

Equations~\eqref{results:Mo} -- \eqref{deltasmall12:1} are the main results of this work. In the next section we present the results of calculations of Fermi polaron longitudinal-transverse splittings and discuss the obtained expressions in more detail.

\begin{figure*}[t]
    \centering
    \includegraphics[width=0.8\textwidth]{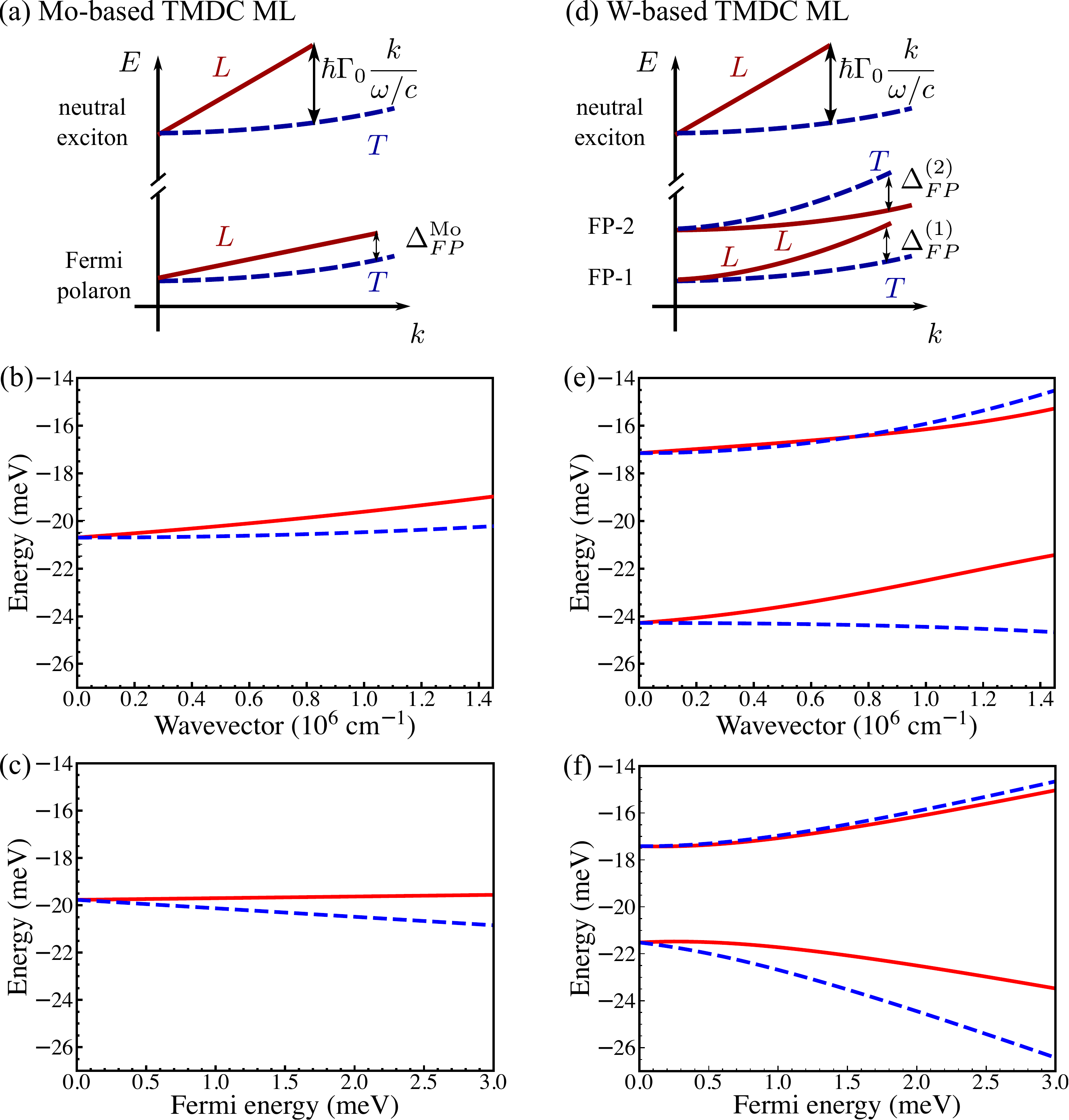}
    \caption{\textbf{Longitudinal-transverse splitting of Fermi polarons.} (a-c) Mo-based monolayer, (d-f) W-based monolayer. Panels (a,d) present a schematics of the level structure of attractive Fermi polarons based on the analysis in Sec.~\ref{subsec:LT}. Panels (b,e) show the splittings of Fermi polaron radiative doublets as functions of the wavevector at a fixed Fermi energy $E_F=2$~meV. Panels (c,f) show the splittings of Fermi polaron radiative doublets as functions of the Fermi energy at a fixed wavevector $k = 10^6$cm$^{-1}$.
    Solid and dashed lines show the energies of the longitudinal ($L$) and transverse ($T$) polarized states. Calculations are performed after Eqs.~\eqref{results:Mo} and~\eqref{results:W:1} and full expressions for the splittings, Eqs.~\eqref{FP:symm:split}. The origin of energy is the neutral exciton energy at $k=0$, $E_F=0$. The remaining parameters used in calculations are: $M_e = M_h = 0.4m_0$, $E_T = 20$meV, $\Delta = 4$meV, $E_g-E_X = 1.7$eV, $\hbar\Gamma_0 = 0.5$meV}.
    \label{fig:splittings}
\end{figure*}

\section{Discussion}\label{sec:disc}

Figure~\ref{fig:splittings} shows the fine structure splittings of Fermi polarons in two-dimensional transition metal dichalcogenides. Panels (a-c) and (d-f) present the results for the Mo-based and W-based ML, respectively. Schematically, the level structure is depicted in Fig.~\ref{fig:splittings}(a,d). Other panels present the numerical results. 

In agreement with simple analytical Eq.~\eqref{results:Mo:1}, for Mo-based ML the splitting between the longitudinal, $L$, and transverse, $T$, Fermi polarons increases linearly as a function of the polaron wavevector $k$, Fig.~\ref{fig:splittings}(b). It  is also a linear function of the Fermi energy, Fig.~\ref{fig:splittings}(c). Here, the only contribution to the LT-splitting is due to the exciton admixture to the correlated state, first term in Eq.~\eqref{Psi:k:+}. 

The situation is more involved in the case of W-based MLs. Particularly, as a function of the wavevector, as shown in Fig.~\ref{fig:splittings}(e), an interplay of $k$-linear and $k^2$ terms in the splitting (first and second terms in Eqs.~\eqref{results:W:1}) results in the non-trivial wavevector dependence of the Fermi polaron LT-splitting. It results in the sign change of the splitting of the upper Fermi polaron doublet. It is because the first, $k$-linear, exciton-related contribution $\propto |\varphi_k|^2$ to the Fermi polaron splitting has an opposite sign compared to the second, $k^2$-contribution related to the intra- and inter-valley polaron mixing by the long-range exchange interaction. The critical value of the wavevector $k_c$ at which the behavior changes from linear to quadratic can be estimated as $k_c \sim \sqrt{\Delta^2 \mu_{eX}/(\hbar^2 E_T)}$.
This second contribution due to the mixing of the inter- and intravalley polarons is similar to the strain induced Fermi polaron splittings in W-based TMDC MLs~\cite{Iakovlev_2023}. In a relevant case of intermediate wavevectors where the mixing of the intra- and intervalley Fermi polarons can be neglected together with the $k$-linear contributions $\propto E_F/E_T$ as compared to the $k^2$ contributions $\propto E_F/\Delta$, we have in agreement with Eqs.~\eqref{results:W:1} and \eqref{Delta:FP:W}

 \begin{equation}
     \label{Delta:FP:W:1}
     \Delta_{FP}^{(1)} = - \Delta_{FP}^{2}= -k^2\frac{3\pi}{8}\frac{\alpha E_F}{\Delta}\sqrt{\frac{\frac{\hbar^2}{2\mu_{eX}}}{E_T}}\frac{c}{\omega}\Gamma_0.
 \end{equation}
Hence, in this case the Fermi polaron splittings are quadratic in the wavevector. Within the range of applicability of Eq.~\eqref{Delta:FP:W:1}, the splitting described by this expression exceeds that of Fermi polarons in Mo-based MLs, Eq.~\eqref{results:Mo:1}. In this case the transverse state of the Fermi polaron with a smaller binding energy [Fermi polaron 2 in Fig.~\ref{fig:splittings}(d)] is above the longitudinal state. At the same time, the longitudinal component of the lower Fermi polaron state 1 is higher in energy compared to its transversal component. Similarly to the case of neutral excitons, both in Mo- and W-based TMDC MLs the splittings of Fermi polarons are proportional to the exciton radiative decay rate $\Gamma_0$. It is because the radiative decay rate controls the long-range exchange interaction strength. Naturally, at $E_F=0$ the fine structure splittings of the Fermi polarons vanish both for the W- and Mo-based MLs because at the absence of doping the Fermi polarons and trions are indistinguishable. 
 
According to the presented calculations, the splittings of the Fermi polarons due to the long-range exchange interaction can be significant and reach meV range for reasonable values of the Fermi polaron wavevector and electron density.

\section{Conclusion}\label{sec:concl}

We have developed a theory of the Fermi polarons (also known as Suris tetrons) energy spectrum fine structure related to the long-range exchange interaction between the electrons and holes. Symmetry analysis shows that, unlike half-integer spin trions, Fermi polarons can have a non-trivial fine structure due to their integer spin nature. Our microscopic calculations demonstrate that, indeed, owing to a correlation of the trion with the Fermi sea hole in a Fermi polaron, these quasiparticles possess the longitudinal-transverse splitting like the neutral excitons. In Mo-based transition metal dichalcogenide monolayers the Fermi polaron longitudinal-transverse splitting is linear in the wavevector. For the W-based structures the dependence of splittings of the wavevector is more complex and can be sign-alternating due to the presence of two states, intra- and intervalley ones, and their mixing by the long-range exchange interaction. 

The longitudinal-transverse splittings of Fermi polarons control the energy spectrum fine structure of these Coulomb-correlated complexes. They are also important for the spin and valley dynamics of Fermi polarons providing efficient depolarization and decoherence similarly to that of the neutral excitons~\cite{maialle93,glazov2014exciton}.  In W-based transition monolayers the off-diagonal terms $\hat{\mathcal H}_{\bm k}^{12}$ can be important for intra- and intervalley Fermi polaron coupling and cross-relaxation, c.f. Ref.~\cite{Zipfel:2020tq}. 

While our calculations were mainly focused on the transition metal dichalcogenide monolayers with resident electrons, similar effects can be observed in $p$-type monolayer semiconductors, van der Waals heterostructures, and in conventional semiconductor quantum wells.

\section*{Acknowledgements}

This work was supported by RSF project No. 23-12-00142.

\bibliographystyle{model1-num-names}





\end{document}